\documentclass[prl,twocolumn,preprintnumbers,amsmath,amssymb,showpacs]{revtex4}
\usepackage{graphicx, epsf,bm,color}
\usepackage{subfigure,amssymb}

\newcommand{\bk}{{\bf k}}
\newcommand{\bq}{{\bf q}}

\begin{document}
\title{\bf Damping of long wavelength collective modes in spinor Bose-Fermi mixtures}
\author{J. H. Pixley}
\affiliation{Condensed Matter Theory Center and Joint Quantum Institute, Department of Physics, University of Maryland, College Park, Maryland 20742- 4111 USA}
\author{Xiaopeng Li} 
\affiliation{Condensed Matter Theory Center and Joint Quantum Institute, Department of Physics, University of Maryland, College Park, Maryland 20742- 4111 USA}
\author{S. Das Sarma}
\affiliation{Condensed Matter Theory Center and Joint Quantum Institute, Department of Physics, University of Maryland, College Park, Maryland 20742- 4111 USA}

\date{\today}
\begin{abstract}
Using an effective field theory we describe the low energy bosonic excitations in a three dimensional ultra-cold mixture of spin-1 bosons and spin-1/2 fermions. We establish an interesting fermionic excitation induced generic damping of the usual undamped long wavelength bosonic collective Goldstone modes.  Two states with bosons forming either a ferromagnetic or polar superfluid are studied.  
The linear dispersion of the bosonic Bogoliubov excitations is preserved with a renormalized sound velocity. 
For the polar superfluid we find both gapless modes (density and spin) are damped, whereas in the ferromagnetic superfluid we find the  density (spin) mode is (not) damped.
We find that this holds for any mixture of bosons and fermions that are coupled through at least a density-density interaction. In addition, we predict the existence of the Kohn anomaly in the bosonic excitation spectrum of Bose-Fermi mixtures.  We discuss the implications of our many-body interaction results for experiments on Bose-Fermi mixtures.
\end{abstract}


\maketitle

The interplay of bosons and fermions is ubiquitous throughout physics, ranging from the interaction of light (i.e. photons) and matter (i.e. electrons) to the behavior of a simple metal in the ionic lattice background.  In solid state physics
the interaction of electrons with slowly moving phonons provides the necessary attractive electron-electron interaction to form Cooper pairs leading to superconductivity~\cite{Bardeen-1957}.  
While on the other hand the effect of the electron Fermi surface on bosons appears naturally in phonon excitations in the form of the Kohn anomaly~\cite{Kohn-1959}.
Another well-know example of the solid state manifestation of fermion-boson interaction is the polaron formation in ionic insulators.
In the description of itinerant quantum phase transitions~\cite{Hertz-1976,Millis-1993} (relevant for 
various strongly correlated materials~\cite{Ishiguro-1990,Hewson-book,Armitage-2010,Kamihara-2008})
the fermions are always coupled to a bosonic collective mode reflecting the ordering of the underlying Fermi gas.  A coupled fermion-boson interacting many-body system is thus a fundamental paradigm in condensed matter physics leading to a large number of interesting phenomena.
  
Coupled fermion-boson systems have also become of interest recently in ultra-cold atomic systems.
In the context of cold atoms, a variety of Bose-Fermi mixtures, e.g., $^6$Li-$^7$Li~\cite{2001_Hulet_LiLi_Science,2001_Schreck_PRL}, $^{40}$K-$^{87}$Rb~\cite{2002_Roati_KRb_PRL,Zirbel-2008,2008_Ni_Ye_Science,2010_Ospelkaus_Ye_Science,2012_Jin_Ye_PRL,Gunter-2006,2013_Bloom_Jin_BF_PRL}, $^6$Li-$^{133}$Cs~\cite{2013_Cheng_LiCs_PRAR,2014_Cheng_LiCs_PRL},  $^6$Li-$^{174}$Yb~\cite{2014_Takahashi_LiYb_arXiv}, have been prepared in experiments for different purposes such as implementing sympathetic cooling~\cite{2001_Hulet_LiLi_Science,2001_Schreck_PRL,2002_Roati_KRb_PRL},  studying molecule formation~\cite{2010_Ospelkaus_Ye_Science}, engineering dipolar  quantum simulators~\cite{2012_Jin_Ye_PRL}, exploring few-body physics~\cite{2013_Bloom_Jin_BF_PRL,2013_Cheng_LiCs_PRAR,2014_Cheng_LiCs_PRL} or looking for interesting collective excitations~\cite{2014_Takahashi_LiYb_arXiv}.  In a recent experiment~\cite{2014_Takahashi_LiYb_arXiv} of particular relevance to our theory to be presented in the current work, laser spectroscopy was used to study the effect of the fermions on the bosonic excitation spectra of $^6$Li-$^{174}$Yb atomic mixtures.

In the absence of fermions, the low energy excitations in Bose-Einstein condensates are well described by Bogoliubov theory and  it is well understood that the Bogoliubov quasiparticles (BQs) are damped through higher order interactions in the form of Beliaev~\cite{Beliaev-1958,Hohenberg-1965,Popov-1972,Liu-1997,Giorgini-1998,Natu-2013} and Landau~\cite{Pitaevskii-1997} damping. Due to destructive quantum interference Landau and Beliaev process are suppressed~\cite{Ozeri-2005} at low momentum and low energies, thus making the long wavelength collective mode a well defined undamped bosonic excitation; in fact, the long-wavelength damping goes as $q^5$ vanishing rapidly as 
$q\rightarrow 0.$ 
An important question of fundamental interest, which has also become relevant in view of recent experiments~\cite{2014_Takahashi_LiYb_arXiv,ferrier-2014}, is, however, still open in spite of extensive theoretical activity, namely, how the fermion-boson interaction (specifically the existence of the Fermi surface) affects the bosonic excitation spectrum in a Bose-Fermi cold atom mixture.  We address this important question in the current work using field theoretic techniques, finding a generic fermion-induced damping of the long wavelength bosonic collective modes.

We start with a microscopic Hamiltonian for a Bose-Fermi mixture, and derive a low energy effective field theory based on a controlled perturbative expansion, from which  the low energy bosonic excitations and damping effects are obtained.   A mixture of a spin-$1$ Bose gas and a spin-$1/2$ Fermi gas is considered and spin SU(2) symmetry is assumed for both theoretical simplicity and relevance to experimental systems in the absence of magnetic fields. 
  We show quite generally that the linearly dispersive BQs of a bosonic superfluid interacting with fermions become damped due to Fermi surface effects, with a damping rate
\begin{equation}
\gamma(\bq)/\hbar = \mathcal{D} |\bq|,
\label{eqn:damp1}
\end{equation}
where $\mathcal{D}$ is dependent on the microscopic details of the system. Moreover, at large momentum ($q \approx k_F$) we show that the Kohn anomaly appears as a kink in the BQ exception spectrum.
The linear momentum dependence in Eq. (\ref{eqn:damp1}) is drastically different from pure boson systems~\cite{Beliaev-1958,Hohenberg-1965,Popov-1972,Liu-1997,Giorgini-1998,Pitaevskii-1997,Ozeri-2005} or 
Bose-Fermi mixtures with a Fermi surface that is gapped out~\cite{ferrier-2014,2014_Zhai_BF_arXiv}. 
On the other hand, the functional form of BQ linear dispersion at low energy $v_s |{\bf q}|$ remains intact, with a renormalized sound velocity $v_s$. 
For spin wave excitations, we find damping in a polar (P) ground state whereas for a ferromagnetic (FM) state the spin waves are found to be undamped.

We begin with the model fermion-boson interacting Hamiltonian, $ H = \int d^3 r \left(\mathcal{H}_B + \mathcal{H}_F + \mathcal{H}_{BF}\right)$, with 
\begin{eqnarray}
\mathcal{H}_B &=& \frac{\hbar^2}{2m_B}\nabla \Phi_a^{\dag}\nabla \Phi_a + \frac{U_0}{2}\Phi_a^{\dag}\Phi_{a'}^{\dag}\Phi_{a'}\Phi_{a} 
\\
&+& \frac{U_2}{2}\Phi_a^{\dag}\Phi_{a'}^{\dag}{\bf T}_{ab}\cdot{\bf T}_{a'b'}\Phi_{b'}\Phi_{b},
\\
\mathcal{H}_F &=& \frac{\hbar^2}{2m_F}\nabla c_{\sigma}^{\dag}\nabla c_{\sigma} ,
\\
\mathcal{H}_{BF} &=&  U_{BF}\Phi_a^{\dag}\Phi_a c_{\sigma}^{\dag}c_{\sigma}+J\Phi_a^{\dag}{\bf T}_{ab}\Phi_b\cdot c_{\alpha}^{\dag}\frac{\bm{\sigma}_{\alpha\beta}}{2}c_{\beta},
\label{eqn:ham}
\end{eqnarray}
where repeated indices are summed over, ${\bf T}$ denotes a vector of spin-$1$ matrices, and $\bm{\sigma}$ denotes a vector of Pauli matrices.  The operator $\Phi_a$ destroys a boson in the $m_z=a$ state and $c_{\sigma}$ destroys a fermion with spin $\sigma$.  The interactions $U_0$ and $U_2$ are related to the scattering lengths of each hyperfine state of the bosons through $U_0 = (g^B_0 + 2g^B_2)/3$ and $U_2 = (g^B_2-g^B_0)/3$, with $g_f = 4\pi \hbar^2 a_f/m_B$ for a scattering length $a_f$ in hyperfine state $f$~\cite{Ho-1998}.  Whereas the Bose-Fermi interactions $U_{BF}$ and $J$ are given by $U_{BF} = (g^{BF}_{1/2}+2g^{BF}_{3/2})/3$ and $J=2(g^{BF}_{3/2}-g^{BF}_{1/2})/3$, with $g_{F_{\mathrm{tot}}}^{BF}=2\pi\hbar^2 a_{F_{\mathrm{tot}}}/m_{BF}$ as the s-wave scattering length with total spin $F_{\mathrm{tot}}$ and $m_{BF}$ the reduced mass~\cite{Xu-2014}.
We focus on a repulsive density-density Bose-Fermi interaction $U_{BF}>0$ and  ferromagnetic spin-spin interaction $J<0$. We assume the fermions to be non-interacting, which is valid  for bare repulsive interactions because they are strongly irrelevant in the low energy limit~\cite{1994_Shankar_RG_RMP,1992_Polchinski_arXiv}.

In the absence of the Fermi gas, the spin-$1$ Bose gas can become~\cite{Ho-1998,Ohmi-1998} either a FM superfluid for $U_2<0$ or a P superfluid for $U_2>0$, and we consider both situations theoretically. In the FM phase, the ground state breaks both the U(1) and SU(2) symmetry of the Hamiltonian and as a result hosts two distinct types of Goldstone modes corresponding to gapless density excitations which are of the BQ form that go as $|\bq|$ and of the ferromagnetic spin wave form that go as $\bq^2$.  The P superfluid also breaks the U(1) and SU(2) symmetry and also hosts two distinct sets of gapless density and spin wave excitations, however in this case they both take on the linear-in-$q$ BQ form.  
 
\begin{figure}
\centering
  \includegraphics[width=0.6\linewidth]{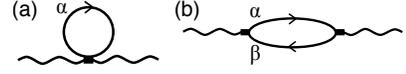}
\caption{
Diagram (a) is referred to as the ``tadpole'' and (b) is the spin dependent particle-hole ``bubble''.
A solid line denotes a fermion Green function with spin $\alpha$, a solid rectangle denotes the Bose-Fermi interaction ($U_{BF}$ or $J$), and the wavy line denotes the bosonic field $\phi_a$ in one of the $m_z=1,0,-1$ states. }
\label{fig:diagrams}
\end{figure}

We are interested in the low energy theory of the system and derive the corresponding effective action for the gas within a path integral framework~\cite{Negele-book}.  The model Hamiltonian corresponds to
an action in the grand canonical ensemble
$
S = \int d\tau \Big(H(\tau) + \int d^3r \big[\Phi_a^{\dag}\partial_{\tau}\Phi_a + c_{\sigma}^{\dag}\partial_{\tau}c_{\sigma} - \mu_B \Phi_a^{\dag}\Phi_a - \mu_Fc_{\sigma}^{\dag}c_{\sigma}\big] \Big)=S_B+S_F+S_{BF}.
$
We have introduced the Bose gas action $S_B$, the Fermi gas action $S_F$, and their mutual interaction $S_{BF}$.  To derive the low energy effective action of the bosonic superfluid we first expand the Bose operators about the mean field ground state $\Phi^T =  \Phi^T_{MF}  + ( \phi_1,  \,\, \phi_0, \,\,  \phi_{-1})$, which together with the mean field value of $\mu_B$ gives rise to an effective action for the $\phi$ degrees of freedom $S_{B,\mathrm{eff}}[\phi]$ (see the Supplemental Material).  As the Fermi gas is non-interacting we integrate them out and determine the effective action of the fluctuations $\phi$ 
 (ignoring any constants)
$
\tilde{S}[\phi] = S_{B,\mathrm{eff}}[\phi] - \mathrm{Tr}\log(1 - \hat{V}(\phi)\hat{G_0}).
$
The trace is over spin, space, and imaginary time indices and we have defined the two by two matrices $\hat{V}(\phi)$ and the Green function of the fermions $\hat{G_0}$, which depend on the superfluid state.  As 
$\hat{V}(\phi)$ is composed entirely of fluctuations of the Bose gas we can expand the logarithm to obtain the low energy effective action to quadratic order
\begin{equation}
S_{\mathrm{eff}}[\phi] = S_{B,\mathrm{eff}}[\phi]+ \mathrm{Tr}( \hat{V}(\phi)\hat{G_0}) +\frac{1}{2}\mathrm{Tr}\left( (\hat{V}(\phi)\hat{G_0})^2\right).
\label{eqn:Seff}
\end{equation} 
The first term after $S_{B,\mathrm{eff}}$ corresponds to the tadpole diagram 
and the second to the spinful particle-hole bubble in Figs. \ref{fig:diagrams} (a) and (b) respectively.

\emph{Ferromagnetic Ground State}:
Focusing on the ferromagnetic case $U_2<0$ ($\langle  \Phi^{\dag}_{MF}{\bf T}\Phi_{MF} \rangle=\rho_0\hat{z}$, where $\rho_0$ is the boson density), the mean field expectation value of the Bose operators can be written as a three component spinor, $\Phi^T_{MF} = (\sqrt{\rho_0}, \,\, 0, \,\, 0)$.  The chemical potential of the bosons at the mean field level is
$
\mu^{FM}_B = g_2^B \rho_0+ J M_z+U_{BF}n_F,
$
where $M_z = (\langle n_{\uparrow} \rangle- \langle n_{\downarrow} \rangle)/2$ and $n_F = \sum_{\sigma} \langle n_{\sigma} \rangle$, with $n_{\sigma}=c_{\sigma}^{\dag}c_{\sigma}$.  Expanding the Bose operators about their mean field ground state results in the fermions seeing an effective magnetic field $\rho_0 J$ in the $\hat{z}$ direction and it breaks the up-down spin symmetry of the Fermi gas.   
This gives rise to the non-interacting Green function,  $\hat{G}_{0\sigma\sigma'}^{-1} = -\delta_{\sigma,\sigma'}(\partial_{\tau} -\hbar^2/(2m_F)\nabla^2-\tilde{\mu}_F + J\rho_0\sigma/2)$, where $\sigma = \pm 1$ for $\uparrow$ and $\downarrow$, and we have shifted the fermionic chemical potential $\tilde{\mu}_F = \mu_F - \rho_0 U_{BF}$. 
The matrix $\hat{V}_{FM}$ is given by
\begin{eqnarray}
& &\hat{V}_{FM}(\phi)=\left(J(\sqrt{\rho_0/2}\phi_0^{\dag}+\frac{1}{2}S_{\phi}^-)\sigma_+ +\mathrm{h.c} \right)
\\
&+& U_{BF}(\sqrt{\rho_0}[\phi_1^{\dag}+\phi_1]+n_{\phi})\bm{1} +\frac{J}{2}(\sqrt{\rho_0}[\phi_1^{\dag}+\phi_1]+S_{\phi}^z)\sigma_z, \nonumber 
\end{eqnarray}
where $\bm{1}$ denotes the identity matrix, $\sigma_i$ are the Pauli matrices, $\sigma_{\pm} = (\sigma_x\pm i \sigma_y)/2$, and we have introduced $n_{\phi}=\sum_a \phi_a^{\dag}\phi_a$ and $S^{\alpha}_{\phi} = \phi_a^{\dag}T^{\alpha}_{ab}\phi_b$.

\begin{figure}
\begin{minipage}[t]{.25\textwidth}
  \includegraphics[width=0.9\linewidth]{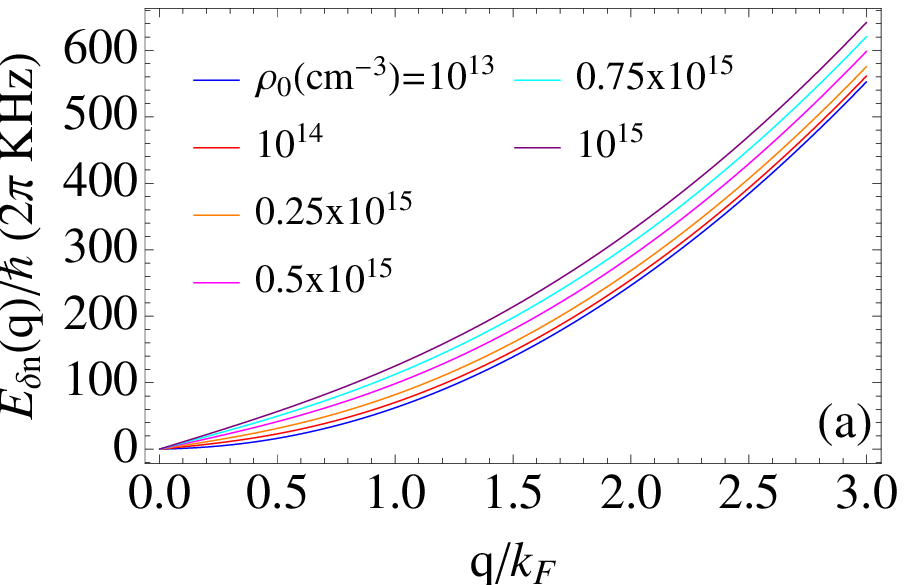}
\end{minipage}%
\begin{minipage}[t]{.25\textwidth}
  \includegraphics[width=0.9\linewidth]{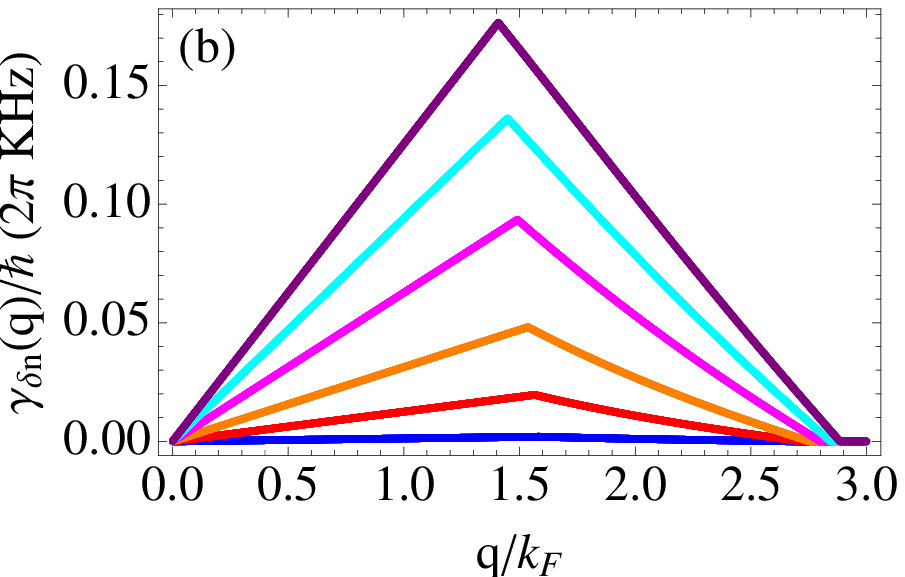}
\end{minipage}
\begin{minipage}[t]{.25\textwidth}
  \includegraphics[width=0.9\linewidth]{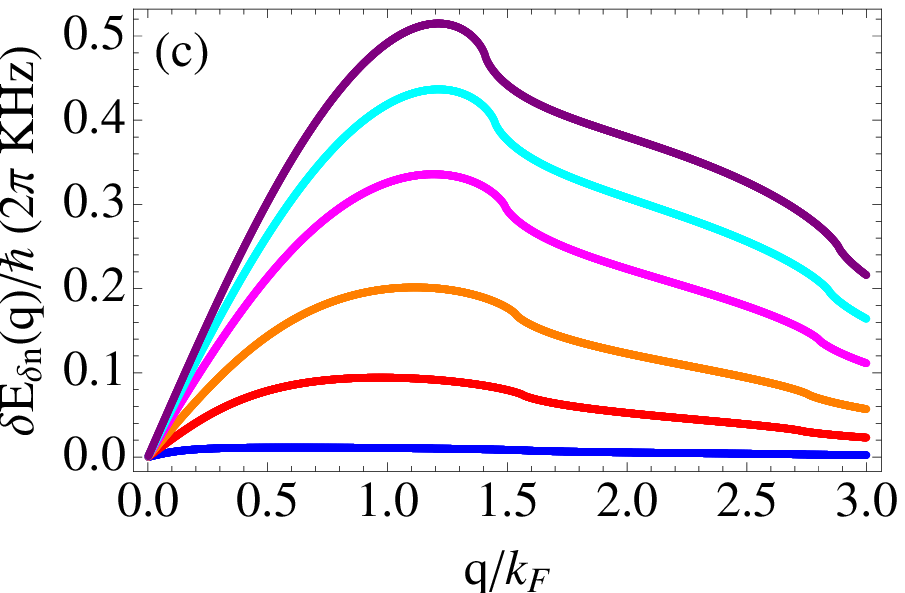}
\end{minipage}%
\begin{minipage}[t]{.25\textwidth}
  \includegraphics[width=0.9\linewidth]{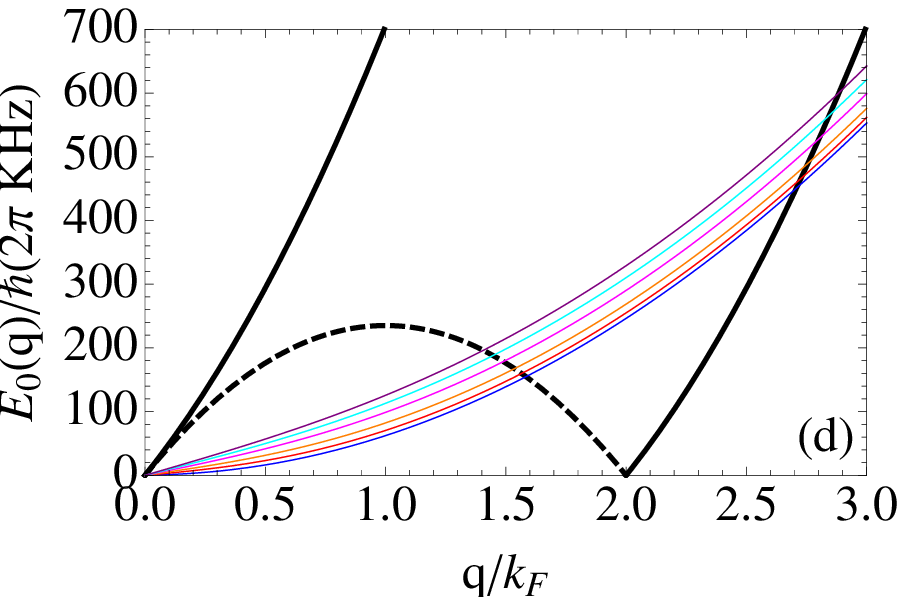}
\end{minipage}
\caption{ Energy spectrum and damping rate of collective modes from solving
Eq.\ (\ref{eqn:pole}) for various bosonic densities $\rho_0$. 
In this plot the exact Lindhard function~\cite{Lindhard-1954} is included focusing on the density mode ($\delta n$) for the P case with $^{23}$Na for the bosons, the Fermi gas it taken to be $^6$Li with a density of $n_F = 10^{13}\mathrm{cm}^{-3}$,  and focusing on $U_{BF}=U_0$, and $J=U_2$. 
(a) The dispersion $E_{\delta n}(\bq)$, (b) the damping rate $\gamma_{\delta n}(\bq)$, (c) the difference in dispersions with and without the Fermi gas $\delta E_{\delta n}(\bq) \equiv E_{\delta n,0}(\bq) - E_{\delta n}(\bq)$, displaying the Kohn anomaly near $q/k_F\sim 1.5$, and (d) the Bose gas dispersion in the absence of the Fermi gas $E_{\delta n,0}(\bq)=\sqrt{v_{s,\delta n,0}^2\bq^2+b_0^2\bq^4}$, all as a function of $q/k_F$. In (d) the black solid lines bound the region of the particle-hole excitation continuum and the black dashed line marks where the imaginary part of the Lindhard function changes its form.}  
 \label{fig:damp2}
\end{figure}

Evaluating the trace for the particle-hole bubble leads to $\chi^0_{\alpha\beta}(q) = \int dk \,G_{0\alpha}(k+q)G_{0\beta}(k)$, we are using the shorthand notation $q\equiv (i\nu_n,\bq)$ $[k\equiv (i\omega_n,\bk)]$  for bosonic [fermionic] Matsubara frequency, 
and the integral $\int dq \equiv \int \frac{dq^3}{(2\pi)^3}\frac{1}{\beta}\sum_n$.  The Green function in momentum space is $G_{0\sigma}(\bk, i\omega_n)^{-1} = i\omega_n - \epsilon_{\bk \sigma}+\tilde{\mu}_F$, with a spin dependent dispersion $\epsilon_{\bk \sigma} = \hbar^2 \bk^2/(2m_F)+J\rho_0\sigma/2$.
For the spin diagonal case $\alpha=\beta$ this reduces to the well known Lindhard function~\cite{Lindhard-1954}, which at sufficiently low temperature in the low energy limit, $|\bq|/k_{F\sigma}\ll1$ and $\nu_n/(\hbar v_{F\sigma}|\bq|)\ll 1$ (where $v_{F\sigma}$ is the Fermi velocity of the spin $\sigma$ electrons),  becomes
\begin{equation}
\chi^0_{\sigma\sigma}(\bq,i\nu_n) = - \eta_{\sigma}\left(1- \frac{\pi}{2}\frac{|\nu_n|}{\hbar v_{F\sigma}|\bq|} - \frac{1}{3}\left(\frac{|\bq|}{2k_{F\sigma}}\right)^2 \right)
\label{eqn:Linfun}
\end{equation}
and we have defined the density of states for spin $\sigma$, $\eta_{\sigma}=m_Fk_{F\sigma}/(2\pi^2\hbar^2)$.  We stress the approximation $\nu_n/(\hbar v_{F\sigma}|\bq|)\ll 1$ is consistent while considering Bose-Fermi mixtures with $v_{s,0}\ll v_{F\sigma}$ where $v_{s,0}$ is the sound velocity of the BQ excitations in the absence of the Fermi gas. Physically, this ensures that the BQs excite a sufficient amount of particle-hole pairs, by passing through their excitation continuum as shown in Fig.~\ref{fig:damp2} (d). For the case $\sigma\neq \sigma'$, in the low frequency, low momentum limit we can always treat $\hbar^2|\bq|^2/(2m_F) \ll |J|\rho_0$, $\hbar v_{F\sigma}|\bq| \ll |J|\rho_0$, and $|\nu_n| \ll |J|\rho_0$.  Within this approximation we find
\begin{eqnarray}
& &\chi^0_{\sigma \bar{\sigma}}(\bq, i\nu_n)  = \frac{1}{\rho_0 J}(\langle n_{\bar{\sigma}} \rangle - \langle n_{\sigma} \rangle)\left(\bar{\sigma} -\frac{i\nu_n}{\rho_0 J} \right)
\\
&+&\Big[\frac{  \hbar^2n_{F}}{2m_F}+\frac{\bar{\sigma}}{5\rho_0 J}\left((\hbar v_{F\bar{\sigma}})^2\langle n_{\bar{\sigma}} \rangle - (\hbar v_{F\sigma})^2\langle n_{\sigma} \rangle \right)\Big]\frac{|\bq|^2}{(\rho_0J)^2}.
\nonumber
\end{eqnarray}

With all of these results in hand we can now determine the effective action to quadratic order.  Writing the action in separate parts we have
$
S^{FM}_{\mathrm{eff}}[\phi] = \int dq \left( \mathcal{L}_{\phi_1}+\mathcal{L}_{\phi_0}+\mathcal{L}_{\phi_{-1}}\right),
$
where
\begin{eqnarray}
&& \mathcal{L}_{\phi_1} =  \phi_1^{\dag}(q)\left(-i\nu_n + b_0\bq^2\right)\phi_1(q)
\label{eqn:S1}
\\
 &+&(\Delta^{FM}_1 +\mathcal{A}^{FM}_1\frac{|\nu_n|}{|\bq|}  + \mathcal{B}^{FM}_1|\bq|^2)\frac{1}{2}\left|\phi_1^{\dag}(q)+\phi_1(-q)\right|^2,
\nonumber
\end{eqnarray}
$ \mathcal{L}_{\phi_0}= (-\mathcal{A}^{FM}_0 i\nu_n + \mathcal{B}^{FM}_0|\bq|^2)\phi_0^{\dag}(q)\phi_0(q),$
and
$ \mathcal{L}_{\phi_{-1}}=\left(\mathcal{M}-i\nu_n +  b_0|\bq|^2\right)\phi_{-1}^{\dag}(q)\phi_{-1}(q),$
where $b_0 = \hbar^2/2m_B$.
In order to simplify the presentation we have introduced the constants $ \Delta_1^{FM},\mathcal{A}^{FM}_1,\mathcal{B}^{FM}_1, \mathcal{A}^{FM}_0,$ and $\mathcal{B}^{FM}_0$, whose functional form is not particularly relevant for the present discussion and
are explicitly given in the Supplemental Material.  Now that we have the Green function for each spin state from the effective action, we can determine the excitation of each mode from the poles~\cite{Negele-book}.  A few remarks are in order: The $\phi_{-1}$ mode is gapped with a value $\mathcal{M} =2(\rho_0|U_2|+M_z|J|)$. In contrast, the spin wave excitations that correspond to $\phi_0$, are given by $\omega_{sw}^{FM} = \hbar^2\bq^2/(2m_s^*) $ where the coefficient is altered from $\hbar^2/(2m_B)$, and the spin excitations acquire a renormalized effective mass $\hbar^2/(2m_s^*) =  \mathcal{B}^{FM}_0/\mathcal{A}^{FM}_0$, which is \emph{not} a function of $U_{BF}$. Therefore, the renormalization of $\mathcal{M}$ and $m_{s}^*$ is due entirely to the paramagnon excitations of the Fermi gas through the coupling $J$.  The density modes ($\phi_1+\phi_1^{\dag}$) acquire damping through the additional contribution of $|\nu_n|/|\bq|$ which arises due to the particle-hole excitations of the Fermi gas (see Eq. (\ref{eqn:damping}) below).  It is useful to note that the constants $\mathcal{A}^{FM}_1$ and $\mathcal{B}^{FM}_1$ are both quadratic functions of $U_{BF}$ and $J$. 


\emph{Polar Ground State}:
We now discuss the polar ground state of the bosons when $U_2>0$, $(\langle \Phi^{\dag}_{MF}{\bf T}\Phi_{MF} \rangle = 0)$,  
with a mean field expectation 
 $\Phi_{MF}^T = (0, \sqrt{\rho_0}, 0)$.  Now, the up down spin symmetry of the Fermi gas remains intact.  The chemical potential for the bosons takes the form
$
\mu^P_B =U_0\rho_0 +U_{BF}n_F.
$
The non-interacting Green function is 
now spin independent, which becomes $\hat{G}_{0\sigma\sigma'}^{-1} = -\delta_{\sigma,\sigma'}(\partial_{\tau} -\hbar^2/(2m_B)\nabla^2-\tilde{\mu}_F)$.  The matrix $\hat{V}_P$ is defined as
\begin{eqnarray}
& &\hat{V}_P(\phi)=U_{BF}(\sqrt{\rho_0}[\phi_0^{\dag}+\phi_0]+n_{\phi})\bm{1} +JS_{\phi}^z\sigma_z/2
\nonumber
\\
&+& \left(J(\sqrt{\rho_0/2}[\phi_{-1}^{\dag}+\phi_{1}]+\frac{1}{2}S_{\phi}^-)\sigma_+  +\mathrm{h.c.}\right).
\end{eqnarray}
At this point it is useful to compare this with the FM case.  From the effective bosonic action $S_{B,\mathrm{eff}}[\phi]$ in the P case we know that the density mode is related to $\delta n(q)^{\dag}=(\phi_0^{\dag}(q),\phi_0(-q))$ while the spin excitations are given by $\delta S(q)^{\dag}=(\phi_1^{\dag}(q), \phi_{-1}(-q))$.  As a result, it is quite natural that the density excitations only couple through $U_{BF}$ and the spin excitations through $J$.  This is quite different from the case of the FM superfluid where the BQs are a combination of density and spin along the $z$ direction.

Following the same steps as before, we compute the trace 
in Eq. (\ref{eqn:Seff}), only now the equal spin particle-hole bubble in Eq. (\ref{eqn:Linfun}) comes into play
 Due to the spin symmetry the Fermi gas parameters are now $\sigma$ independent.
The effective action to quadratic order for the polar case is given by,
$
S^P_{\mathrm{eff}}[\phi] = \int dq \left( \frac{1}{2}\mathcal{L}_{\delta n}+\mathcal{L}_{\delta S}\right),
$
where $\mathcal{L}_{\delta n}=\delta n(q)^{\dag} G_{\delta n}(q)^{-1} \delta n(q)$ and $\mathcal{L}_{\delta S}=\delta S(q)^{\dag} G_{\delta S}(q)^{-1} \delta S(q)$.  The bosonic Green functions can be written in terms of the free bosonic Green function and the self energy using Dyson's equation
$
G_{a}(i\nu_n,\bq)^{-1}=G_{a}^0(i\nu_n,\bq)^{-1}-\Sigma_{ a}(i\nu_n,\bq),
$
which are two by two matrices
$
G_{a}^0(i\nu_n,\bq)^{-1}= 
-i\nu_n \sigma_z + b_0\bq^2 \bm{1} 
$
and
\begin{eqnarray}
\Sigma_{a}(i\nu_n,\bq)= -\left(\Delta^P_a +\mathcal{A}^{P}_a\frac{|\nu_n|}{|\bq|}  + \mathcal{B}^{P}_a|\bq|^2\right)\left(\bm{1}+\sigma_x \right).
 \label{eqn:Sigma}
\end{eqnarray}  
We give the explicit forms of $\Delta^P_a, \mathcal{A}^{P}_a$, and $\mathcal{B}^{P}_a$ in the Supplemental Material.  In this case, the $\delta n$ constants 
are quadratic functions of $U_{BF}$ and do not depend on $J$ while the $\delta S$ constants are quadratic functions of $J$ and are independent of $U_{BF}$.  Interestingly, in the polar case we find both BQ modes to be damped through the appearance of $|\nu_n|/|\bq|$.
We remark that the density mode for the FM case in Eq. (\ref{eqn:S1}) can also be converted to the above form.

\emph{BQ Excitation Spectrum and Damping}:
We have derived the effective field theory for both the ferromagnetic and polar bosonic ground states interacting with a spin-$1/2$ Fermi gas.  
To diagonalize the effective action using the Bogoliubov transformation we find $u_{\bk}$ and $v_{\bk}$ have to be functions of $\nu_n$.  
To determine the dispersion and damping rate we find the poles of the bosonic Green function  $G_a$ (with $a=\phi_1,\delta n,\delta S$) ~\cite{Negele-book}, after analytically continuing the Matsubara frequency to real frequency $i\nu_n \rightarrow \omega + i 0^+$, we solve the algebraic equation 
\begin{equation}
\mathrm{det}\,G_a(\omega,\bq)^{-1} = 0,
\label{eqn:pole}
\end{equation}
to find the poles at $\hbar\omega = E({\bq}) - i\gamma(\bq)$, where $\gamma(\bq)/\hbar \ge 0$ is the damping rate.
We obtain two of our main results
\begin{eqnarray}
\gamma_{a}(\bq) =b_0\mathcal{A}_a|\bq|,   \,\,\,\,\,\,\,\,\, E_a(\bq)=\sqrt{v_{s,a}^2\bq^2+b_a^2 \bq^4},
\label{eqn:damping}
\end{eqnarray}
(see Fig.\ \ref{fig:damp2}) and we find the Bogoliubov form remains
with the renormalized parameters
$
v_{s,a}^2  =b_0(2\Delta_a-b_0\mathcal{A}_a^2),
$
and
$
b_{a}^2 = b_0\left(b_0+2\mathcal{B}_a \right).
$
The low energy theory can also be extended to finite temperature, which yields $\gamma_a(\bq,T)\propto [1+\tanh(\tilde{\mu}_F/k_B T)]|\bq|$ and the linear momentum dependence survives.
Solving Eq.\ (\ref{eqn:pole}) using the exact Lindhard function~\cite{Lindhard-1954} gives results indistinguishable from the analytical results as shown in 
Fig.\ \ref{fig:damp2} (b) up to 
$q\sim O(k_F)$. 
Thus, all BQ modes present in a spin-$1$ Bose gases coupled to a Fermi gas become damped at low energies and are thus no longer true collective modes even at long wavelengths!

Several comments about our main results shown in Eq.\ (\ref{eqn:damping}) and
Fig.\ \ref{fig:damp2}
are in order: (i) Our results are valid to all orders in the Bose-Fermi coupling provided $v_{s,0}\ll v_F$ since all higher-order vertex corrections in the self-energy are negligible by virtue of Migdal's theorem~\cite{Migdal-1958}. (ii) The fermion-induced bosonic damping being linear in $q$ strongly dominates any intrinsic bosonic Beliaev damping at long wavelength~\cite{Liu-1997}.
(iii) Eq.\ (\ref{eqn:damping}) implies that the bosonic collective mode frequency vanishes without becoming overdamped if the damping becomes comparable or larger than the mode energy itself since both the frequency and the damping go linear in wave number.
(iv) At large wave numbers, $q>k_F$,
 $\gamma_a(\bq) \rightarrow 0$ because the imaginary part of the Lindhard function~\cite{Lindhard-1954} itself vanishes by energy-momentum conservation and the BQs can no longer excite particle-hole pairs [see Fig.\ \ref{fig:damp2} (d)]
(v) The spinor structure of the problem is not essential to find this Fermi gas induced damping, therefore our results are valid for any condensed spin-$S$ Bose gas coupled to a Fermi gas (either spinful or spinless) through at least a density-density interaction $U_{BF}\hat{n}_B\hat{n}_F$.

The Fermi surface has effectively been imprinted upon the excitation spectrum of the Bose gas in the form of the Kohn anomaly.  As this is a large momentum effect, it is not captured by our low energy theory but comes directly out of the numerical solution using the exact Lindhard function. However, this effect is weak for the atomic mixtures we are considering and is not visible in $E_a(\bq)$ itself, only the difference $\delta E_a(\bq) = E_0(\bq) - E_a(\bq)$, 
displays the anomaly as seen in Figs.\ \ref{fig:damp2} (a) and (c).


Our theoretical predictions can be tested in ultra cold Bose-Fermi mixtures with weakly interacting fermions, through two-photon Bragg spectroscopy~\cite{Ozeri-2005} that couples to the bosonic species. 
We expect the damping will give rise to a broadened Bragg line, i.e. an intrinsic line width, in the presence of fermions that will depend explicitly on energy-momentum as shown in Eq. (\ref{eqn:damping}). Due to the bosonic damping persisting to sufficiently low momentum, we also expect that the presence of the Fermi gas can dephase the bosonic superfluid, which has possibly already been observed in Ref.\ \onlinecite{Gunter-2006}.

In the presence of a shallow harmonic trap as in most atomic experiments, 
we expect our results to be largely unaffected for large clouds of atoms. Even though the finite size of the trap gives an infrared cut off to the theory, this energy scale is sufficiently small such that linear-in-$q$ BQ dispersion can still be observed in experiments~\cite{Steinhauer-2002}. In this limit, the physics is well described by the Thomas-Fermi and local density approximations, where the BQs will still be linear-in-$q$ at low momentum (with a renormalized sound velocity)~\cite{Ho-1998,Zambelli-2000,Steinhauer-2002} and the Fermi gas will acquire a spatially dependent Fermi velocity. 
In spite of this, close to the center of the trap, the form of particle-hole excitations will remain the same with a renormalized $v_F$.
However, we would like to mention that these inhomogenous effects can in principle be removed by considering a uniform potential trap, which has recently been achieved, where the spatial inhomogeneity is minimized~\cite{2013_Hadzibabic_BoxTrap}.

We have studied a spin-$1$ Bose gas coupled to a spin-$1/2$ Fermi gas in three dimensions.  We have shown BQ excitations are damped at long wavelength as a result of particle-hole excitations of the Fermi gas while the functional form of the BQ dispersion is unchanged.  
We have argued this phenomenon should apply to Bose-Fermi mixtures in general independent of the spin structure of either species.

\emph{Acknowledgements}
We would like to thank Stefan Natu and Brian J. DeSalvo for useful discussions. This work is supported by JQI-NSF-PFC and ARO-MURI.

\bibliography{sBFm}

\newpage
\onecolumngrid
\setcounter{figure}{0}
\makeatletter
\renewcommand{\thefigure}{S\@arabic\c@figure}
\setcounter{equation}{0} \makeatletter
\renewcommand \theequation{S\@arabic\c@equation}

\section*{\Large{Supplemental Material}}
In the supplemental material we give the explicit expressions for $S_{B,\mathrm{eff}}$ and the equations defining the constants that we have introduced in the main text in terms of the parameters of the model.
For the ferromagnetic case, the effective bosonic action including the bosonic chemical potential is
\begin{eqnarray}
S^{FM}_{B,\mathrm{eff}}[\phi] &=& \int d\tau d^3r \left(\sum_a\phi_a^{\dag}\partial_{\tau}\phi_a + \frac{\hbar^2}{2m_B}\sum_a\nabla \phi_a^{\dag}\nabla \phi_a +g_2^B\rho_0 \frac{1}{2}(\phi_1^{\dag}+\phi_1) (\phi_1^{\dag}+\phi_1) + 2\rho_0|U_2|\phi_{-1}^{\dag}\phi_{-1} \right)
\nonumber
\\
&-& \int d\tau d^3r \left((JM_z+U_{BF}n_F)\left[\sqrt{\rho_0}(\phi_1^{\dag}+\phi_1) + \sum_a \phi_a^{\dag}\phi_a \right]\right),
\end{eqnarray}
whereas for the polar case, the effective bosonic action takes the form
\begin{eqnarray}
S^P_{B,\mathrm{eff}}[\phi] &=& \int d\tau d^3r \left(\sum_a\phi_a^{\dag}\partial_{\tau}\phi_a + \frac{\hbar^2}{2m_B}\sum_a\nabla \phi_a^{\dag}\nabla \phi_a 
+\frac{U_0\rho_0}{2} (\phi_0^{\dag}+\phi_0)(\phi_0^{\dag}+\phi_0) + U_2\rho_0 (\phi_1^{\dag}+\phi_{-1})(\phi_{-1}^{\dag}+\phi_1) \right)
\nonumber
\\
&-&\int d\tau d^3r U_{BF}n_F\left(\sqrt{\rho_0}(\phi_0^{\dag}+\phi_0)+\sum_a\phi_a^{\dag}\phi_a \right).
\end{eqnarray}

To simplify the presentation of the main text we have defined the following constants for the ferromagnetic case
\begin{eqnarray}
\Delta_1^{FM} &=& \rho_0 g_2^B- \rho_0\left(U_{BF}^2+\frac{1}{4}J^2\right)\sum_{\sigma}\eta_{\sigma}-\rho_0U_{BF}J\sum_{\sigma}\sigma\eta_{\sigma},
\\
\mathcal{A}_1^{FM} &=& \rho_0\left(U_{BF}^2+\frac{1}{4}J^2\right)\frac{\pi}{2}\sum_{\sigma}\frac{\eta_{\sigma}}{\hbar v_{F\sigma}}+\rho_0U_{BF}J\frac{\pi}{2}\sum_{\sigma}\sigma\frac{\eta_{\sigma}}{\hbar v_{F\sigma}},
\\
\mathcal{B}_1^{FM}  &=& \rho_0\left(U_{BF}^2+\frac{1}{4}J^2\right)\frac{1}{12}\sum_{\sigma}\frac{\eta_{\sigma}}{k_{F\sigma}^2}+\rho_0U_{BF}J\frac{1}{12}\sum_{\sigma}\sigma\frac{\eta_{\sigma}}{k_{F\sigma}^2},
\\
\mathcal{A}_0^{FM}&=& 1+\frac{M_z}{\rho_0},
\\
\mathcal{B}_0^{FM}&=&\frac{\hbar^2}{2m_B}+\frac{n_{F}}{\rho_0}\frac{  \hbar^2}{2m_F}+\frac{1}{10\rho_0^2J}\sum_{\sigma}\sigma(\hbar v_{F\sigma})^2n_{\sigma} .
\end{eqnarray}

For the polar case we have introduced the constants
\begin{eqnarray}
\Delta_{\delta S}^P &=& \rho_0U_2 - \frac{1}{2}\rho_0\eta J^2,
\\
\mathcal{A}_{\delta S}^P &=& \frac{\pi\rho_0\eta J^2}{4\hbar v_F},
\\
\mathcal{B}_{\delta S}^P &=& \frac{\rho_0\eta J^2}{24 k_F^2},
\\
\Delta_{\delta n}^P &=& \rho_0U_0 - 2\rho_0\eta U_{BF}^2,
\\
\mathcal{A}_{\delta n}^P &=& \frac{\pi\rho_0\eta U_{BF}^2}{\hbar v_F},
\\
\mathcal{B}_{\delta n}^P &=& \frac{\rho_0\eta U_{BF}^2}{6 k_F^2}.
\end{eqnarray}

\end{document}